**Title:** Lack of Ecological Context Can Create the Illusion of Social Success in *Dictyostelium discoideum*

**Authors:** Ricardo Martínez-García & Corina E. Tarnita*

**Affiliation:** Department of Ecology and Evolutionary Biology, Princeton University, Princeton NJ 08544, USA

*Corresponding author email: ctarnita@princeton.edu

**Abstract:**

Studies of cooperation in microbes often focus on one fitness component, with little information about or attention to the ecological context, and this can lead to paradoxical results. The life cycle of the social amoeba *Dictyostelium discoideum* includes a multicellular stage in which not necessarily clonal amoebae aggregate upon starvation to form a possibly chimeric (genetically heterogeneous) fruiting body made of dead stalk and spores. The lab-measured reproductive skew in the spores of chimeras indicates strong social antagonism; this should result in low genotypic diversity, which is inconsistent with observations from nature. Two studies have suggested that this inconsistency stems from the one-dimensional assessment of fitness (spore production) and that the solution lies in tradeoffs between multiple traits, e.g.: spore size versus viability; and staying vegetative versus becoming dormant. We theoretically explore different tradeoff-implementing mechanisms and provide a unifying ecological framework in which the two tradeoffs above, as well as novel ones, arise collectively in response to characteristics of the environment. We find that spore production comes at the expense of vegetative cell production, time to development, and, depending on the experimental setup, spore size and viability. Furthermore, we find that all existing experimental results regarding chimeric mixes can be qualitatively recapitulated without needing to invoke social interactions, which allows for simple resolutions to previously paradoxical results. We conclude that the complexities of life histories, including social behavior and multicellularity, can only be understood in the appropriate multidimensional ecological context.

**Author summary**

Cooperation, in which individuals provide benefits to others at a cost to themselves, has been studied extensively, and mechanisms have been proposed for its persistence in the face of free-riders, individuals that obtain the benefits of the social contract without paying any cost. Often however, especially in microbes, these studies focus on one fitness component, with little information about or attention to the ecological context in which species live, which can lead to paradoxical findings. Here we discuss such an example in the slime mold *Dictyostelium discoideum*, and provide a unifying ecological framework in which multiple life history tradeoffs arise collectively in response to characteristics of the environment. We show that this multidimensionality can resolve existing inconsistencies and explain all the experimental results that have been traditionally attributed to the existence of complex interactions among strains, such as social cheating. We conclude that the complexities of social behavior can only be understood in the appropriate ecological context.

**Introduction**

The cellular slime mold, *Dictyostelium discoideum* is one of the most studied examples of cooperation and altruism in microbes. Upon starvation, solitary amoebae aggregate with neighbors to form a multicellular fruiting body made of stalk and spores. The spores are resistant and will germinate upon encountering favorable conditions while the stalk cells die during stalk development [1-4]. In the process of aggregation these amoebae do not exclude non-kin; consequently, chimeras (multicellular fruiting bodies consisting of at least two genotypes) have been observed both in the lab and in nature [5-8]. These chimeras are functional: the multiple genotypes participate both in stalk formation and in spore production (although not necessarily in equal measures, a phenomenon known as reproductive skew [6]). Therefore, this is an ideal organism for the study of potentially very costly social behavior [3]. Studies to date have found significant reproductive skew in *D. discoideum* chimeras [9,10] and in a variety of other cellular slime molds [11]. These findings point towards a decrease in species-wide genetic diversity that is inconsistent with the immense diversity and coexistence observed among strains in nature [5,8,9,11].

Recent studies [12,13] have suggested that this inconsistency arises due to the one-dimensional assessment of *D. discoideum* fitness, which is equated to spore contribution, and proposed that life-history tradeoffs between non-social traits lead to multiple fitness components. One proposed tradeoff is between spore number and viability and was determined empirically: genotypes that are overrepresented in spores also made smaller and less viable spores when grown clonally [13]. A second proposed tradeoff is between staying vegetative and becoming a spore [12]. In *D. discoideum* not all cells aggregate to become multicellular; experiments have shown that these cells are viable and theoretical approaches suggested that they could be part of a bet-hedging strategy in uncertain environments: vegetative cells have a high chance of death if the starvation period is long, but, if food does return to the environment, they have a head start against spores that need time to germinate [12,14]. Given this tradeoff, genotypes that are overrepresented in spores could simply be those that have been selected to leave fewer vegetative cells behind [12]. Such tradeoffs can, under certain ecological conditions, resolve the coexistence inconsistency [12].

While [12-14] convincingly argue that *D. discoideum* fitness has multiple components, there is no immediate link between the two different tradeoffs they propose, nor is there a theoretical framework in which to assess (a) how many such tradeoffs are likely to occur, or (b) by what mechanisms they play out. Here, we set out to provide such an eco-evolutionary framework in which we study selection on non-social life-history traits in *D. discoideum*, explore the possible tradeoffs that can arise, and make testable predictions for future empirical work. While the framework is not all-encompassing, it provides a theoretical starting point on which additional ecological and life-history knowledge can be built. Within our theoretical framework, both tradeoffs identified in [12-14], as well as novel ones and relationships between them, arise collectively and synergistically. We recapitulate existing experimental results without needing to invoke social interactions and we propose improved experimental designs and measures of *D. discoideum* fitness that capture the variety of tradeoffs in diverse environments.

More broadly, these findings affect our understanding of *D. discoideum* social behavior and multicellularity and they resonate with recent studies emphasizing the challenges associated with assessing cooperation and free-riding in microbes in the absence of the ecological context under which the trait deemed cooperative has evolved and is maintained [12-19]. We showcase some of the complex outcomes (e.g. bet-hedging, coexistence) that ecological context and selection on non-social traits can produce in the absence of social interactions. Many of these outcomes have been theoretically shown before when studied independently (i.e. focus on one trait or one environmental characteristic, e.g. see [20-22]) but here we identify powerful synergistic interactions that can not be predicted from one-dimensional analyses. Finally, these results also contribute towards clarifying the misinterpretations that can arise from sociobiological investigations into microbe behavior that are not grounded in an ecological understanding.

### Eco-evolutionary framework

The model we employ is a generalization of [12] to study more broadly life-history tradeoffs in *D. discoideum* in response to environmental stressors. Since we are interested in nonsocial traits, we study a well-mixed population in which we assume that there are no social interactions between genotypes and in which we do not model the spatial aggregation process. This model therefore ignores interactions between genotypes that may arise during the developmental process. Although *D. discoideum* and other cellular slime molds are likely to be found in spatially structured environments where movement in the vegetative state is limited, a first well-mixed approach is necessary to tease apart the effects of spatial structure from effects arising from simple life history tradeoffs in a well-mixed setting. In this socially neutral, well-mixed context, aggregation occurs randomly (with anyone in the population) and within chimeric aggregates there are no interactions.

*Environment.* Here we focus on starvation and we study both deterministic environments – in which food recovery is certain and the starvation times (time between onset of starvation and the next food pulse) are always of the same length, $T$, and stochastic environments – in which food recovery is variable and uncertain and the starvation times are drawn from an exponential distribution with average $\lambda_T$ (Fig. 1A). We assume that all cells in a given environment compete for the available resources.

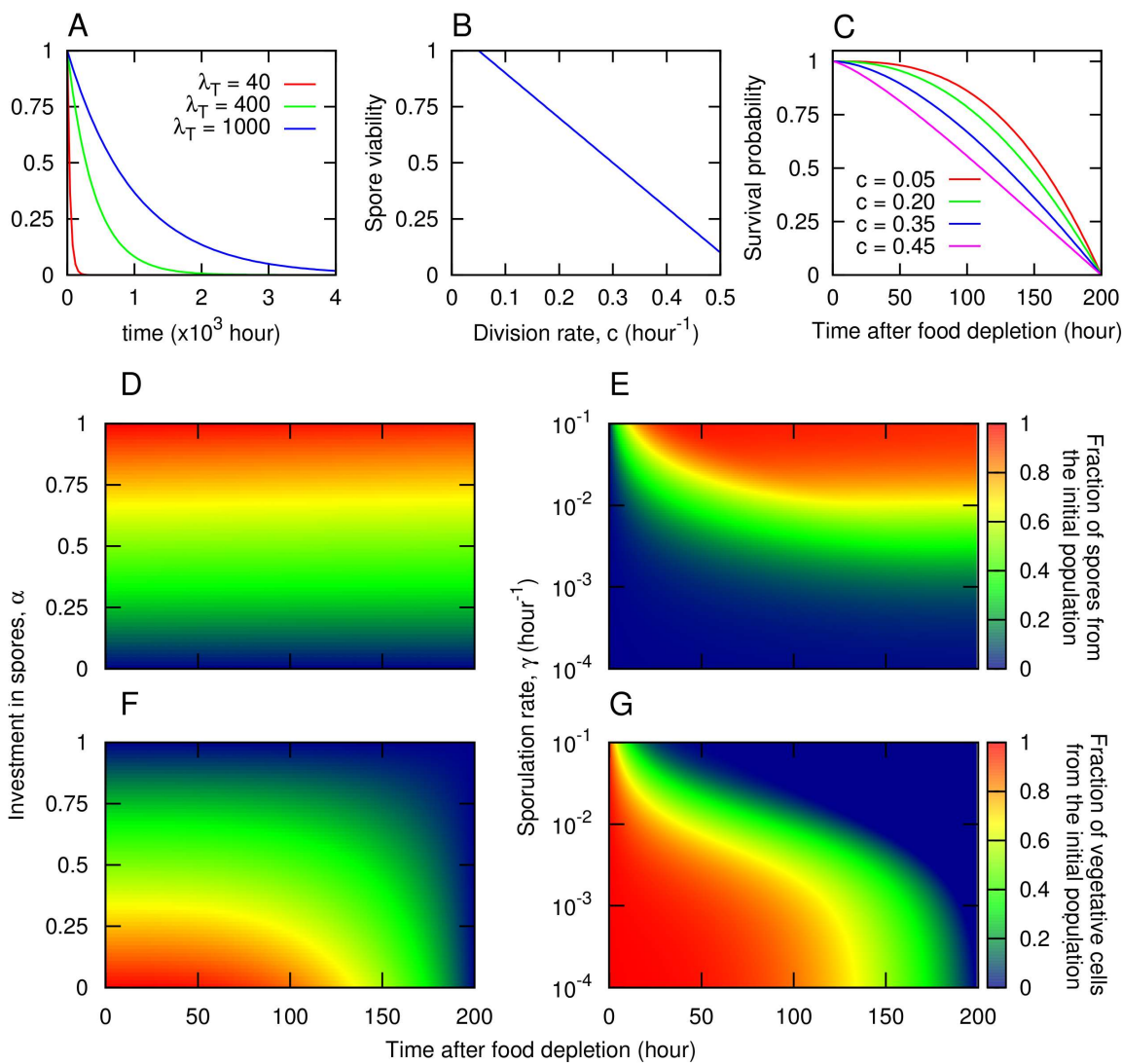

**Figure 1. Framework assumptions.** A) Food recovery in a given environment is stochastic following an exponential distribution (unnormalized for clarity) with mean $\lambda_T$. B) Spore viability as a function of the division rate, $c$. C) Survival curves for starving vegetative cells with different division rates, $c$. D, F) Discrete spore-making mechanism: genotypes, $\alpha$, make spores (fraction $\alpha$) and loners (fraction $1 - \alpha$) instantaneously after starvation. E, G) Continuous spore-making mechanism: genotypes, $\gamma$, make spores at rate $\gamma$. D, E) Fraction of the population at the beginning of the starvation period that has turned into spores at time $t$ after starvation. F, G) Fraction of the population at the beginning of the starvation period that remains vegetative at time $t$ after starvation. Spores also die at a very small rate $\delta$, but this effect is imperceptible at short times scales as in this figure. For D—G we used the survival curve corresponding to $c = 0.15$. Other parameters are specified in Table S1.

*Life-history traits*. Genotypes can evolve in a two-dimensional life-history trait space determined by their doubling time, $c$, and their response to starvation. The first trait, the doubling time, $c$, captures the tradeoff between cells that divide fast and are able to use up common resources but pay the cost of a decreased survival (for vegetative cells) or viability (for spores, [13]), and cells that divide slowly thereby potentially losing out resources to faster dividing strains but at the same time having an increased survival or viability (Fig. 1B,C). The second trait, the response to starvation, encapsulates a genotype's decision to commit resources to sporulation or to remaining vegetative, and captures the tradeoffs between resistant spores that are able to survive in harsh conditions but are slow to germinate when the conditions improve, and non-resistant vegetative cells that eventually die in harsh conditions but are able to immediately start eating and dividing if food returns.

*Sporulation tradeoff implementing mechanisms*. For the latter tradeoff we study two possible mechanisms via which a genotype's reaction to starvation could be implemented. The first mechanism we propose is a discrete one (e.g. stochastic switching), whereby the population splits instantaneously upon starvation into fixed fractions of spores and non-aggregators. Then the characteristic of a genotype is the fraction of individuals that aggregate upon starvation, $\alpha$ (Fig. 1D,F). Thus, if $\alpha = 0$, a monoculture of genotype $\alpha$ does not undergo aggregation; if $\alpha = 1$, a monoculture only produces spores and leaves no vegetative cells behind. Intermediate $\alpha$ values represent a mixed strategy, where some cells sporulate and others do not. The second mechanism we propose is a continuous one, whereby vegetative cells turn into spores continuously at rate $\gamma$, which is then the characteristic of the genotype. The higher the rate $\gamma$, the faster the cells become spores (Fig. 1E,G). While in the discrete approach only a certain fraction of cells can become spores and the others remain vegetative, in the continuous approach every cell has the potential to become a spore but some take longer than others (Fig. 2A). Therefore this latter approach also allows the study of development time, defined as the time it takes for all spores to get formed. With this definition, the lower the sporulation rate $\gamma$ the longer it takes a genotype to complete development (Fig. 2B).

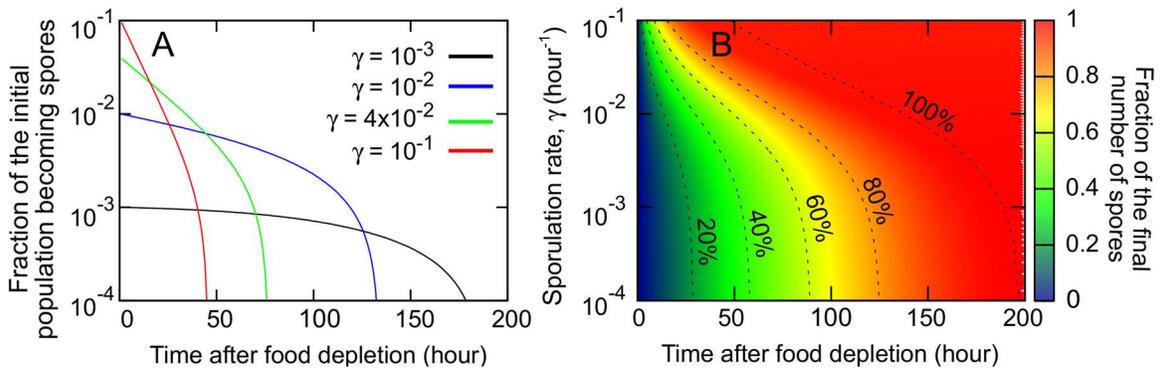

**Figure 2. Phenotypic diversity and development time.** A) After starvation, cells of a given genotype turn into spores at rate $\gamma$. Genotypes with higher $\gamma$ make most of their spores immediately; genotypes with lower $\gamma$ take longer to produce the spores. B) Spore formation velocity for a given $\gamma$ (development time).

*Model: growth phase.* The dynamics of the free-living amoebae is given by a well-mixed model of competition for a resource $R$ between the different genotypes, each given by a pair ($\alpha$, $c$) or ($\gamma$, $c$), depending on the mechanism under investigation (see Methods for the equations). Resources are consumed according to Michaelis-Menten kinetics. We assume that in the growing phase cell death rate is zero. This is a good approximation since most cellular death is caused by starvation. Cells eat and divide as long as resources are available; when resources have been exhausted, cells enter a *starvation phase*, whose length is determined by the environmental conditions and might be constant (deterministic environments) or sorted from an exponential distribution (stochastic environments). In this phase the population splits between spores and vegetative (non-aggregating) cells by one of the two mechanisms above – discrete or continuous. The mathematics of these two scenarios is detailed in the Methods section.

During the starvation phase, spores and vegetative cells die at different rates. Spores die at a constant rate while vegetative cells die according to survival curves that depend on the cell size, which we assume to be inversely correlated with the doubling time or division rate $c$ (Fig. 1C and see Methods section for the formula describing the family of survival curves). Experimental results have demonstrated that in the absence of food non-aggregating cells survive by consuming their own organelles and cytoplasmic resources via autophagy [23]. This results in a low mortality rate at short times. Once they consume all the intracellular resources the mortality rate increases and finally all vegetative cells die after a maximum lifetime. We propose that, by analogy with life history tradeoffs in other species [20], strains that reproduce faster have smaller cells, with fewer cytoplasmic resources and consequently have a higher mortality rate at short times. Reproducing faster has therefore an inherent cost in terms of a less mature and more vulnerable offspring.

The starvation phase is followed by a new *growth phase* induced by the arrival of the next food pulse. The non-aggregators surviving the starvation phase start reproducing immediately according to Eq.(1) in Methods. The spores however start their germination process lasting for several hours, and only after its completion they are able to start eating and dividing again. Different genotypes will have different germination success due to their characteristic spore viability, which is dependent on spore size [13]. The spore viability was chosen to be a linear function of the reproduction rate $c$ (inversely proportional to cell size), so that strains with the fastest reproduction have the lowest spore viability and, at the other extreme, strains with the slowest reproduction have the highest spore viability (Fig. 1B). Finally, only a fraction of the germinating spores result in viable cells to account for aggregating cells that are lost as dead stalk. These growth—starvation cycles continue indefinitely.

**Results/Discussion**

*Winning genotypes (deterministic environments).* Consistent with previous results [12,14], deterministic environments always select for pure strategies: environments where food recovers faster select for all vegetative cells ($\alpha = 0$, respectively $\gamma = 0$) and environments where food recovers slower select for all spores ($\alpha = 1$, respectively $\gamma = 10^{-1}$; although $\gamma$ can be infinitely large, for computational feasibility we impose a large but finite upper limit on $\gamma$, in this limit, at the end of development, more than 90% of the starving cells become spores) (Fig. 3A,C). This holds regardless of whether we use a discrete or a continuous spore-forming mechanism. The switch between these two regions (all-vegetative cells versus all-spores) takes place at starvation times close to the maximum lifespan of a vegetative cell in the absence of food. In both regions, the lower the starvation time, the lower the doubling time $c$. In the first region (selection for all-vegetative cells), due to the very short starvation periods, the whole population increases over time so that each new growth period starts with a higher initial cell density. Since we assume that the food pulse is always the same size, then an increasing population finishes the food faster and thus reduces the length of the growing periods. The shorter the growing periods, the less advantage for the fast reproducing strains. Hence, for very short starvation periods, fast reproducing strains have an initial growth advantage but, over many growth-starvation cycles, they are outcompeted by slow reproducing ones as growth benefits diminish and are outweighed by the higher survival costs incurred during the starvation phase (Fig. S1A,B). As the length of the starvation periods increases, faster reproducing strains start to win (Fig. 3A,C, S1A,C). In the second region (selection for all-spores), we find the same increasing trend in $c$ as the length of the starvation period increases (Fig. 3A,C). This is due to the fact that few spores die during short starvation times and therefore, the initial population sizes responding to a new food pulse are large and able to consume the food quickly. This results in short growing periods, and consequently strains that reproduce faster do not get enough divisions during a growth cycle to overcome the cost of having a lower spore viability (Fig. S1A,D).

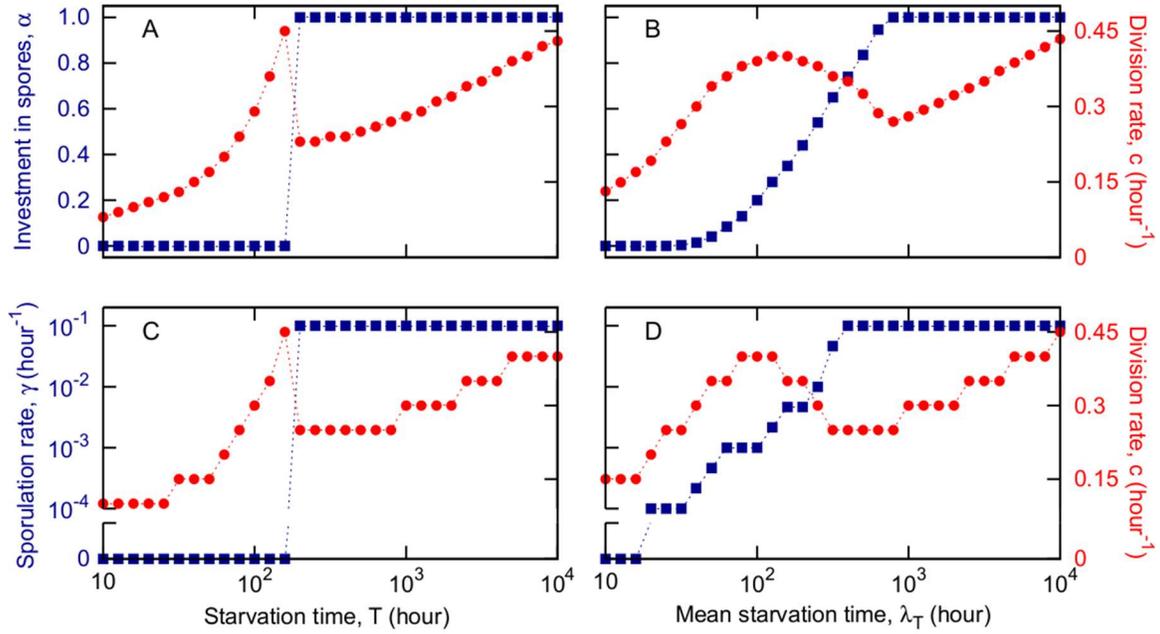

**Figure 3. Winning genotypes in deterministic and stochastic environments under discrete and continuous sporulation.** In all panels blue squares represent the investment in spores (sporulation rate) and red circles represent the division rate. A) Discrete spore-making and deterministic environments. Winner was determined as the most abundant out of 4141 genotypes (101 values of $\alpha$ and 41 values of $c$) after a single realization of $10^8$ hours. B) Discrete spore-making and stochastic environments. Results averaged over 20 independent simulation runs starting with 44772 genotypes (1092 values of $\alpha$ and 41 values of $c$). C) Continuous spore-making and deterministic environments. Winner determined from a single realization starting with 99 genotypes (9 values of $c$ and 11 values of $\gamma$). D) Continuous spore-making and stochastic environments. Simulations set up as in C but with exponentially distributed starvation times; averages taken over 20 independent realizations. Parameters as in Table S1.

*Winning genotypes (stochastic environments).* Consistent with previous results [12,14] as well as extensive theoretical work on bet-hedging (e.g. [21,24]), stochastic environments always select for one winning strategy that is pure at the extremes – all-vegetative cells for very fast and all-spores for very slow environments – but mixed otherwise (Fig. 3B,D). This holds regardless of whether we use a discrete or a continuous spore-forming mechanism. In this case the doubling times follow similar trends to the deterministic case for fast and slow environments but follow the opposite trend for intermediate environments (Fig. 3B,D).

Henceforth, we will call intermediate environments precisely those for which the reproduction rate of the winning genotype decreases with increasing length of starvation; environments to the left of this region will be called fast and environments to the right of this region will be called slow. The intuition behind the increasing trend in fast and slow environments is the same as for the deterministic case (Fig. S2A,B,D). For intermediate environments, the decreasing trend in division rate of the winning genotype occurs when the starvation periods are long enough to be costly for the survival of vegetative cells but they are still short so that spores incur little mortality leading to short growth periods during which fast reproducing cells do not derive enough benefit to offset the lower viability of their spores. (Fig. S2C). For realistic values of the spore death rate (not too large), these qualitative trends are robust with respect to parameter choice both in the stochastic and in the deterministic case (Fig. S3,S4). The results are independent of the initial conditions (Fig. S5A) but they are affected by the saturation behavior assumed for the Michaelis-Menten dynamics (Fig. S5B). Finally, the two genotypic dimensions are synergistic such that selection in the two dimensional space of spore investment ($\alpha$ or $\gamma$) and division time ($c$) leads to different winning strategies than those obtained by varying only one of the traits and keeping the other fixed (Fig. S6).

Overall, despite the multidimensionality of fitness and the several tradeoffs between the life history traits, we found that each environment, whether deterministic or stochastic, selects for only one winning genotype. Consistent with the predictions of [12] coexistence can be achieved in the current model if we incorporate spatial heterogeneity and weak-to-moderate dispersal between different environments. In the absence of any spatial heterogeneity, multiple genotype coexistence is possible only when strains balance their tradeoffs so that they have identical fitness, as suggested in [13]. Such balancing however reduces the scenarios where coexistence may occur and yields them ungeneric. This is in agreement with classic works in community ecology that show that in the absence of frequency-dependent mechanisms tradeoffs alone do not generally result in coexistence [25].

In what follows we will focus on stochastic environments since they are more likely to capture the realities of microbial lives [26]. When we refer to fast-environment, intermediate-environment or slow-environment genotypes, we mean the winning genotypes from fast, intermediate, respectively slow environments.

*Life history traits and tradeoffs.* To determine tradeoffs between reproduction, survival and viability we used the winning genotypes from each stochastic environment (Fig. 3B for the discrete mechanism and Fig. 3D for the continuous mechanism) and we allowed initial clonal populations of identical size to complete one growth cycle on identical resources and undergo the developmental process triggered by starvation. Thus, unlike in our evolutionary setup above where competing genotypes underwent many successive growth-starvation cycles, for the purposes of exploring life history traits and tradeoffs we replicated laboratory setups of one growth-starvation cycle to allow for a comparison with existing data. When sporulation was completed, we evaluated the population size, number of spores and number of vegetative cells for each of these genotypes and related them to each other and to the division time *c*. Additionally, we associated these traits with the investment in spores $\alpha$ (Fig. S7) for the discrete mechanism, and with the development time (which anticorrelates with $\gamma$), for the continuous mechanism (Fig. 4).

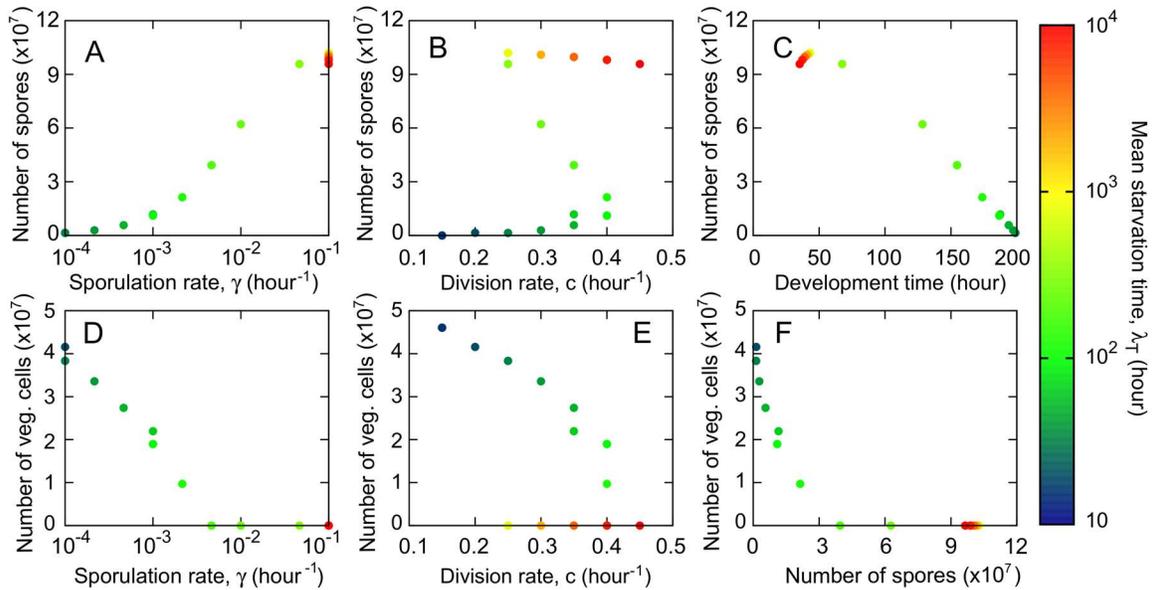

**Figure 4. Life history tradeoffs between non-social traits in genotypes with continuous spore formation.** Top row: spore-related traits are measured after 200 hours of starvation to ensure that all genotypes have completed their development. Number of spores is compared to: A) sporulation rate, B) division rate and C) development time. Bottom row: vegetative cells-related traits are measured after 160 hours of starvation, when the slowest sporulating genotypes have produced 95% of their final number of spores. Number of vegetative cells is compared to D) sporulation rate, E) division rate, and F) number of spores.

In one growth-starvation cycle the spore-forming mechanism does not affect the growth period; furthermore, at the end of the growth period started from identical and monoclonal initial conditions the population size is constant with respect to $c$ (Fig. S7A). This follows from the assumption that resources are used proportional to the division rate and that there is no (or negligible) cell mortality during the growth phase: i.e., strains that divide faster (higher $c$) use more resources per division and therefore reach their carrying capacity sooner, while strains that divide slower (lower $c$) use fewer resources per division and reach their carrying capacity later. However, since all strains start with identical numbers on identical resources, when the resources are exhausted they all reach the same carrying capacity. Thus, when strains grow clonally for only one growth period, higher doubling time does not confer a benefit in terms of increased number of cells; however, it continues to incur a cost during starvation in terms of cell survival. This is an essential difference from the evolutionary setup where mixes of several strains compete for resources during many growth-starvation cycles.

Although population sizes are constant with respect to $c$ at the end of growth regardless of the spore-forming mechanism, during starvation differences emerge between the two mechanisms: whereas for the discrete mechanism sporulation is completed instantaneously at the onset of starvation and therefore all quantities are measured at the end of the growth period, for the continuous mechanism sporulation takes time and during that time vegetative cells of different genotypes sporulate and die at different rates. We will therefore discuss the results for the two mechanisms separately.

For the discrete mechanism, the number of spores is equal to $\alpha P$ and the number of vegetative cells is equal to $(1-\alpha)P$, where $P$ is the population size at the end of the growth phase, and is independent of $\alpha$ and constant with respect to $c$. Thus, the number of spores always increases with $\alpha$ (Fig. S7B) and it has the same behavior with respect to $c$ as does $\alpha$: i.e. the production of spores increases with the division rate $c$ for slow-environment genotypes, decreases with $c$ for intermediate-environment ones and is constant with respect to $c$ for fast-environment genotypes (Fig. S7C). Spores and vegetative cells are anticorrelated, so that a high investment in spores comes at the cost of a low investment in vegetative cells (Fig. S7D). The vegetative cells always decrease with $\alpha$ (Fig. S7E) and with respect to $c$ they have the inverse behavior of $\alpha$: i.e. vegetative-cell production decreases with *c* for fast-environment genotypes and increases for intermediate ones (Fig. S7F). Slow-environment genotypes do not produce vegetative cells.

Continuous spore making genotypes need a finite but non-zero development time to turn a starving population of cells entirely into spores. This time depends on both the sporulation rate $\gamma$ and the division rate $c$: for fixed $c$ it decreases with increasing $\gamma$, since the higher $\gamma$ is, the faster the spores get formed; and for fixed $\gamma$ it decreases with increasing $c$, since the higher the division rate is, the faster the vegetative cells die leaving only the spores. We quantify spore-related traits when all genotypes have completed their sporulation ($t$ = 200 hours after starvation) (Fig. 4 top row). Since at this time there are no more vegetative cells, we quantify vegetative cells-related traits at some intermediate point at which at least some genotypes still have some vegetative cells (Fig. 4 bottom row). For this purpose, here we look at time $t$ = 160 hours but the results hold for any time $t$ prior to complete sporulation.

As explained above, at the end of the growth period all genotypes have the same population size, irrespective of $c$. If $c$ were fixed, the genotypes with higher $\gamma$ would produce spores faster and would therefore have a higher number of spores since fewer of their vegetative cells have enough time to die of starvation. If $\gamma$ were fixed, the genotypes with lower $c$ would have a higher number of spores since their vegetative cells survive longer and therefore are more likely to eventually become spores. However, our winning genotypes vary with both $c$ and $\gamma$ and therefore $\gamma$ and $c$ interact to determine spore number. For intermediate-environment winning genotypes $\gamma$ increases and $c$ decreases simultaneously (Fig. 3B and 3D) so that they enhance each other in their effect on spore number (Fig. 4A,B). For slow-environment winning genotypes $\gamma$ is constant but $c$ increases (Fig. 3B and 3D) and therefore spore number decreases (Fig. 4A,B). For fast-environment winning genotypes both $\gamma$ and $c$ (Fig. 3B and 3D) increase and therefore they affect spore number in opposite directions; their net effect however is dominated by $\gamma$ so that for fast-environment genotypes spore number increases with $\gamma$ and $c$ (Fig. 4A,B). Development time is inversely proportional to the sporulation rate and to the division rate; therefore the number of spores decreases with development time except for the slow-environment genotypes where the increase in $c$ given the fixed (maximal) sporulation rates reverses the trend (Fig. 4C). At an intermediate time at which at least some genotypes still had some vegetative cells we related the number of vegetative cells to division rate, sporulation rate and number of spores. We found that the number of vegetative cells decreases with sporulation rate $\gamma$ and with the division rate $c$ (Fig. 4D,E). Finally, as was the case for the discrete mechanism, the spores and vegetative cells are negatively correlated (Fig. 4F).

It is important to note that these tradeoffs are sensitive to the experimental conditions. All results above hold if monocultures are allowed to undergo a full growth-starvation cycle. If, as has been the case in experimental studies [13], monocultures are grown exponentially in abundant resources and then washed and abruptly starved, we find an overall positive correlation between number of spores and division rate (Fig. S8). This results in a tradeoff between number of spores and cell and spore size that agrees with experimental findings reported in [13].

*Chimeric success*. In this neutral context where individuals do not interact with each other except indirectly via their competition for resources (possible inter-strain interactions during the aggregation and the development of the fruiting body are neglected), chimeric success is only apparent and it is simply a measure of which genotype makes more spores in a mix. In the context of our new understanding of the interplay between life history traits we propose a new measure of chimeric success that emphasizes the importance of both growth and starvation, in contrast to an existing protocol used in *D. discoideum* and other slime molds [10,11,13] that focuses only on the starvation phase. Although chimeras made of several genotypes occur in nature, experimental work has exclusively focused on pairwise mixes and therefore for ease of comparison we will limit our analysis to this case. The difference consists in the setup: for the existing measure, which we will denote $CS_S$, one starts with a 50:50 mix of two starving genotypes and allows them to form spores; for the new measure, which we will denote $CS_{GS}^{X_0}$, one starts an initial population of size $X_0$ composed of a 50:50 mix of two genotypes on an amount of food, $R_0$, and allows them first to grow, then starve naturally, and subsequently form spores. In both setups, in keeping with previous work [9,10,13], the chimeric success of a genotype in a pairwise mix is given by its fraction of the total spores; the overall chimeric success of a genotype is the average over all such pairwise mixes. For mathematical definitions and details see Methods. Although the two setups are different, the measure we propose reduces to the existing one in the limit of high cell relative to resource density – in that case the food is insufficient to support growth and therefore cells starve instantaneously.

The chimeric success is measured upon completion of spore formation. Thus, for the discrete spore formation mechanism chimeric success is measured at the onset of starvation, when all spores are formed instantaneously; for the continuous spore formation mechanism we measure the number of spores when all genotypes will have completed their sporulation (in our case that is *t* = 200 hours after starvation).

For the pairwise comparisons we used the winning genotypes from each environment obtained in Fig. 3B and Fig. 3D respectively and started with 50:50 mixes of low (Fig. 5A, Fig. 6, S9, top row), intermediate (Fig. 5B, Fig. 6, S9, middle row) and high (Fig. 5C, Fig. 6, S9, bottom row) initial cell densities relative to resource magnitude. As before, in what follows, when we refer to fast-environment, intermediate-environment or slow-environment genotypes, we mean the winning genotypes from these respective environments. We want to investigate two aspects. First, we want to explore relative chimeric success between pairs of genotypes (Fig. 5). Second, we want to determine how overall (average) chimeric success correlates with non-social life history traits: division time *c* and implicitly cell survival and spore viability, spore production, development time for the continuous mechanism and investment in spores and vegetative cells for the discrete mechanism (Fig. 6). Because in mixes that are allowed to grow the production of spores and vegetative cells depends on the mixing partner we compare the chimeric success of a genotype not to its production of spores and vegetative cells in monoculture, as determined in the *Life history traits and tradeoffs* section, but to an average production of spores and vegetative cells, where the average is taken over all pairwise mixes.

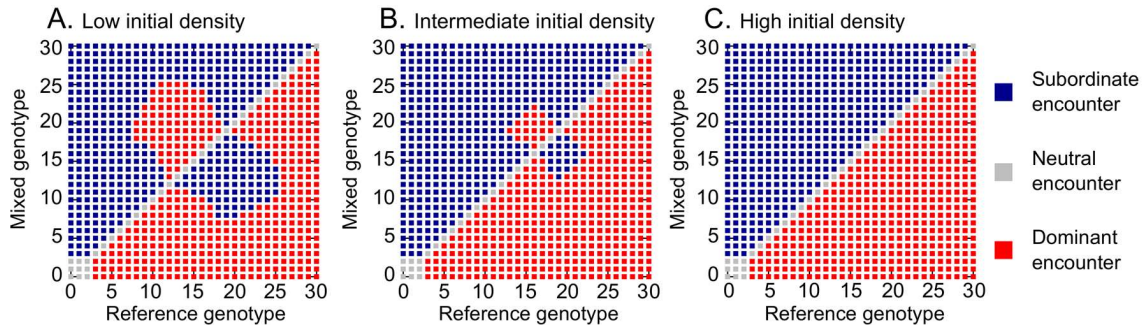

**Figure 5. Relative chimeric success in pairwise mixes.** The winning genotypes from Fig. 3 are mixed in pairs and their relative number of spores measured after a growth-starvation cycle. Genotypes are ordered according to the environment where they evolved in, with 0 corresponding to the environment with $\lambda_T = 10$ hour and 30 to $\lambda_T = 10^4$ hour. A dominant (subordinate) encounter occurs when the reference genotype is overrepresented (underrepresented) in the spore population. In a pairwise encounter, the dominant strain is said to have a higher chimeric success than the subordinate strain. A) Low initial cell:resource density: $10^3$ cells and $R_0 = 10^8$ resources,

B) Intermediate initial cell: resource density: $10^7$ cells and $R_0 = 10^8$ resources, C) High initial cell:resource density: $10^{10}$ cells and $R_0 = 10^8$ resources.

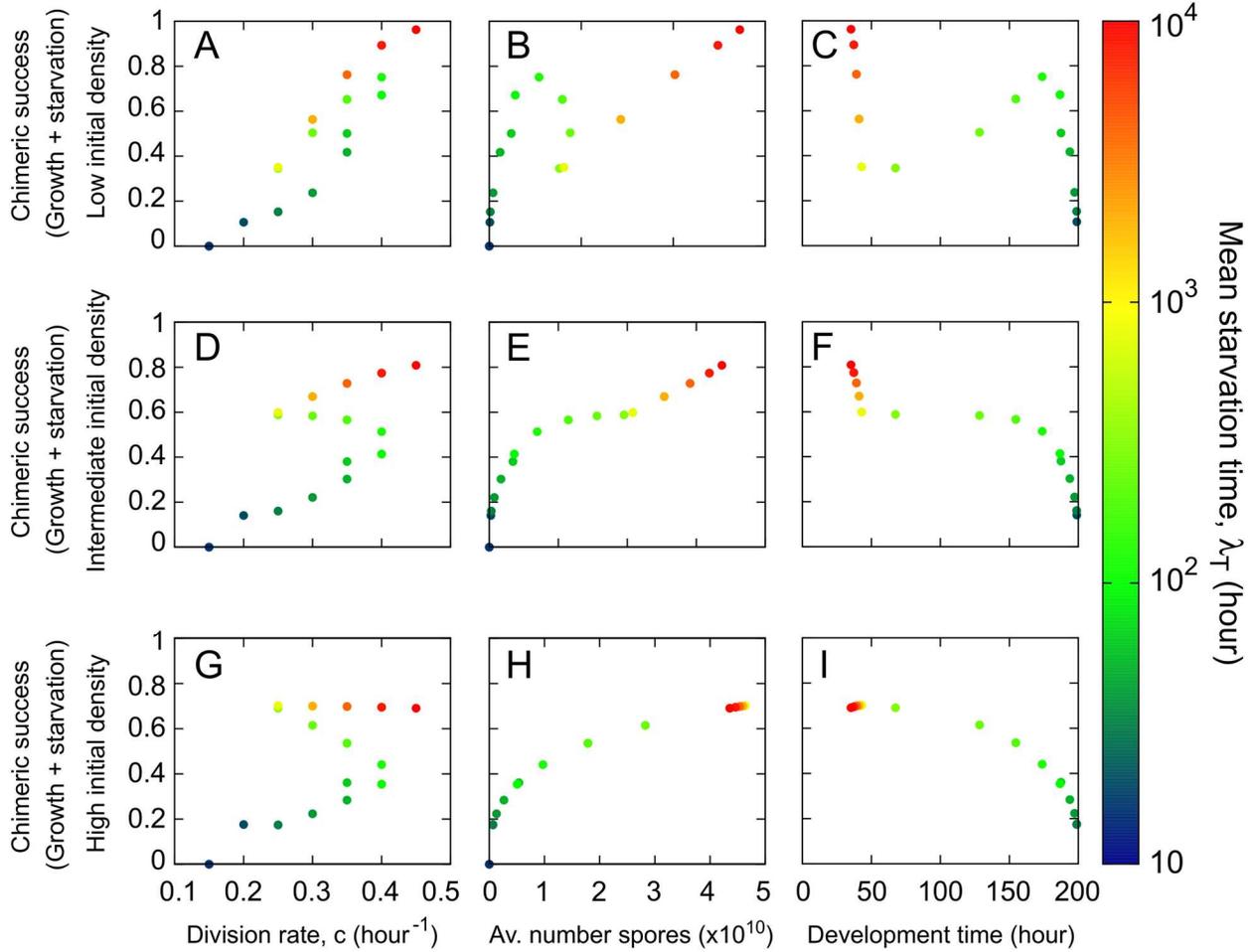

**Figure 6. Apparent life history tradeoffs involving social traits in genotypes with continuous formation of spores.** Top row: average chimeric success starting from a low cell:resource initial density – $10^3$ cells, $R_0 = 10^8$ resources, versus A) division rate, B) average number of spores, C) development time. Center row: average chimeric success starting from an intermediate cell:resource initial density – $10^7$ cells, $R_0 = 10^8$ resources, versus D) division rate, E) average number of spores, F) development time. Bottom row: average chimeric success starting from a high cell:resource initial density – $10^{10}$ cells, $R_0 = 10^8$ resources, versus G) division rate, H) average number of spores, I) development time.

As explained above for clonal growth, in one growth-starvation cycle the spore-forming mechanism does not affect the growth period. Unlike in the case of monocultures, however, where the population size is constant with respect to $c$ at the end of growth, in mixes, a higher division rate $c$ gives the benefit of faster food consumption and increased population size. Although higher $c$ still incurs the cost of higher mortality during starvation, if the growth period is sufficiently long, its potential benefits can outweigh the costs. Since chimeric success is determined upon starvation-induced fruiting body formation following one growth period, what matters is the length of the growth period. This depends on the initial cell density relative to the amount of available food such that the lower the initial cell density relative to the food pulse, the longer the growth period.

Both in the pairwise and in the average scenarios we find that, at low initial densities, chimeric success is mainly determined by growth and therefore its behavior is dominated by the division rate rather than the sporulation rate (or spore investment). As the initial cell density relative to resource magnitude is increased, the two forces – division rate and sporulation rate/spore investment – start to equilibrate and eventually, at high cell density where there is only residual growth, the starvation period dominates and the chimeric success is mostly determined by the sporulation rate/spore investment.

For pairwise comparisons, regardless of the initial density we find that our strains are organized in a linear hierarchy of chimeric dominance, consistent with lab results for *D. discoideum* [9,10] (Fig. 5). However, the ordering of the strains strongly depends on initial densities. For high initial cell density the hierarchy ordering is dominated by the spore investment and we find that the higher the $\gamma$ (respectively $\alpha$), the higher the chimeric success of the genotype (Fig. 5C). As the initial cell density decreases and the division rate starts to play a stronger role, intermediate-environment genotypes for which spore investment is anticorrelated with the division rate separate into two symmetric groups that interchange their places in the hierarchy (Fig. 5A,B): the group with the higher $\gamma$ and lower $c$ gets displaced by the group with the lower $\gamma$ and higher $c$. The slow-environment genotypes continue to top the hierarchy and the fast-environment genotypes continue to be at the bottom of the hierarchy.

Moving on to the overall (average) chimeric success, we find that at low initial densities it is positively correlated with the division rate (Fig. 6A, S9A); it is negatively correlated with development (positively correlated with $\gamma$, respectively $\alpha$) for fast- and slow-environment genotypes, but it is positively correlated with development (negatively correlated with $\gamma$, respectively $\alpha$) for intermediate-environment genotypes where it is most evident that the effect of $c$ dominates the effect of spore investment (Fig. 6C, S9B). Similarly, chimeric success is positively correlated with the average number of spores for fast- and slow-environment genotypes but it is negatively correlated with it for intermediate-environment genotypes for which the negative effect of decreasing $c$ is stronger than the positive effect of increasing $\gamma$ or $\alpha$ (Fig. 6B, S9C). Finally, the behavior of the chimeric success with respect to vegetative cell number is the opposite of the behavior with respect to spores (Fig. S9D).

As the initial cell density relative to resource magnitude is increased, the division rate and sporulation rate/spore investment start to play more equal roles (Fig. 6D-F, S9E-H) and eventually, at high cell density, the sporulation rate/spore investment dominates. Thus, at the high density extreme where growth is only residual and starvation dominates, chimeric success correlates positively with the sporulation rate and the investment in spores (Fig. S9J), and with the number of spores (Fig. 6H, S9K), and it correlates negatively with development time (Fig. 6I) and average number of vegetative cells (Fig. S9L). Chimeric success still correlates positively with division rate for fast-environment genotypes but it correlates negatively with division rate for intermediate-environment genotypes where now the effect of $\gamma$ dominates the effect of $c$ (Fig. 6G, S9I). For slow-environment genotypes where the spore investment ($\alpha$, respectively $\gamma$) is fixed (maximal), the behavior of the chimeric success with respect to the division rate depends on the spore formation mechanism. For the discrete mechanism the correlation with division rate is slightly positive, since a higher division rate implies a higher production of spores. For the continuous mechanism chimeric success is measured after 200 hours of starvation, so cells with a higher division rate also pay a higher survival cost. This results in a smaller population of spores when increasing $c$ and thus a reduced chimeric success.

**Conclusion**

We built a theoretical framework, focused on *D. discoideum* but easily generalizable to other sporulating microbes, to explore the effects of selection on non-social life history traits in variable environments. Within this framework, we were able to qualitatively recapitulate all existing results attributed to apparent social behavior in *D. discoideum* in a model that assumes no social interactions and do not consider any other interaction that may occur during the development of the fruiting body. This highlights the importance of an extended understanding of ecology and life history. In the absence of ecological knowledge – which environment each genotype evolved in, and in the absence of life history knowledge – which life history traits selection acts on and what relationships exist between them, one can incorrectly interpret differences in spore investment as chimeric success and attribute the latter to complex interactions (e.g. social cheating). In the neutral context, previously paradoxical findings do not even arise.

Specifically, we found that the two genotypic traits under selection – division rate and spore investment, interact synergistically to yield tradeoffs between (i) spore investment and vegetative cell investment, such that spore production always comes at a cost of vegetative cell production; and (ii) spore production and development time, such that a higher spore number requires a shorter development time. Furthermore, depending on the experimental protocol, a third tradeoff can be identified between spore production and cell and spore size, survival and viability, such that spore production generally comes at a cost of spore viability, consistent with [13].

In this multi-trait context, measures that attempted to quantify success upon starvation – such as spore number and chimeric success, were revealed to be ill-defined, as we showed starvation and growth to be inextricably linked. We therefore proposed a new and more general measure of chimeric success that accounts for both growth and starvation and found that genotypes are organized in a linear hierarchy based on their pairwise chimeric success, consistent with experimental results [9,10]. Furthermore, the overall chimeric success of a genotype, measured as an average over pairwise comparisons, generally increases with the division rate (i.e. decreases with cell/spore size) consistent with experimental results [13], it increases with the average number of spores produced and with the investment in spores, and it decreases with development time. Finally, we found that the relationship between chimeric success and investment in vegetative cells depends on the experimental conditions. Our results constitute general testable predictions and emphasize the importance of standardized measures and experimental protocols, ideally chosen to most closely approximate natural conditions. Importantly, they call for a more general measure of *D. discoideum* fitness, that would account for the way natural selection acts on the life histories of individuals and their ancestors, in variable environments [27].

Several future directions arise. First, our analysis revealed the importance of mechanisms underlying the non-social traits (e.g. stochastic switching versus phenotypic variation); these need to be explored in future experimental and theoretical work. Second, although we were able to qualitatively capture existing results in a well-mixed, neutral framework, this dismisses neither the importance of spatial structure, which is likely to influence the dynamics at least quantitatively, nor the possibility of social interactions in general. While the linear hierarchies in *D. discoideum* are well aligned with the neutral hypothesis, other slime mold species have exhibited short dominance loops [11], possibly indicative of small subsets of interacting genotypes. Even in *D. discoideum*, there exists at least one extreme lab mutant that has been shown to interact in chimeras, a social parasite that is unable to form its own stalk but uses other genotypes' stalks to support its spores [28]. Although such a mutant has not been found in nature, it has been shown to be very destructive in lab experiments [28] and therefore its effects are worth re-investigating using this new multi-trait understanding of fitness. One hypothesis is that the non-aggregating vegetative cells can act as buffers against the destructive effects of a parasite that can only exploit the social aspects of the behavior. This is consistent with existing results from the theory of cooperation and sociality showing that loners – individuals that do not participate in the social contract – are of primary importance for the evolution of social behavior [29,30]. Third, the life cycle of *D. discoideum* may involve many other tradeoffs to be considered in the future, such as allocating more cells in the spore body at the expense of reducing the dispersal ability by creating a shorter stalk. Furthermore, for simplicity, we have encapsulated here all the ecological variability in the starvation times; however, environmental heterogeneity may also come into play in other ways, e.g. in the pulses of resources, which here we have taken to be constant as a first approach. This could introduce additional tradeoffs arising from strains showing different feeding behavior [31].

Finally, our results show broadly that in fluctuating environments multicellularity and sociality are just part of a set of risk-management strategies. Although our work was motivated by *D. discoideum* in particular and by microbes in general, these results can be extended to other species where similar ecologically-induced tradeoffs between development time, size and social behavior have been identified [32,33].

**Materials and Methods.**

*I. Model:* we develop a well-mixed model for the complete life cycle of *D. discoideum*. It consists in two phases: a growth phase where single amoebae reproduce (and eventually spores germinate) and a starvation phase where cells die and implicitly aggregate through sporulation.

*Growth phase.* The genotypes have abundances $X_{*,c}$, where the $*$ is a place holder for either $\alpha$ or $\gamma$, depending on the mechanism under investigation. Thus

$$\dot{X}_{*,c}(t) = \frac{cR(t)}{R_{1/2} + R(t)} X_{*,c}(t)$$
$$\dot{R}(t) = -\frac{R(t)}{R_{1/2} + R(t)} \sum_{*,c} cX_{*,c}(t) \quad (1)$$

for all genotypes ($*$, $c$). Resources are consumed at a rate governed by a Michaelis-Menten kinetics with saturation constant $R_{1/2}$. When food is abundant, growing rates are related to population doubling times, $T_d$, by $c = \ln 2 / T_d$. We assume that in the growing phase the death rate is zero. This is a good approximation since most cellular death is caused by starvation.

*Starvation phase.* During the starvation phase, spores are dormant and die at a constant rate $\delta$ while vegetative cells die according to their size. We choose a family of survival curves (Fig. 1C) with different short-time decays for every reproductive rate,

$$S_c(t) = \frac{e^{-(\mu t)^{\beta(c)}} - e^{-(\mu T_{sur})^{\beta(c)}}}{1 - e^{-(\mu T_{sur})^{\beta(c)}}} \quad (2)$$

where $\mu$ is the speed at which the death rate of a vegetative cell changes with time since starvation and $\beta(c)$ is a function of the reproductive rate accounting for the cost of the reproduction speed ($\beta > 1$ to ensure a slow decay at short times). Therefore, $S_c(t)$ gives the probability of being alive at time t. Due to the lack of experimental data for the survival curves, we picked $\beta$ satisfying the assumption that it is a decreasing function of the growing rate, reflecting that strains with a lower doubling time have to pay a cost in terms of high short-time death rates (smaller values of $\beta$ result in survival curves that decay faster at short times). For the analysis in the main text we use $\beta(c) = 3.1 - 4c$ which leads to the curves in Fig. 1C. However, we also performed a sensitivity analysis for the choice of $\beta$ (Fig. S3, S4), which showed that the shorter the interval in $\beta$ (more similar survival curves) the faster reproduction is favored in a given environment. This is the expected result, since having similar survival reduces the relative cost of reproducing fast while keeping fixed the benefits.

We compare two mechanisms of spore formation, a discrete one and a continuous one. (1) Discrete (e.g. stochastic switching). In this case spores and non-aggregators are determined instantaneously after starvation onset. Thus, during the rest of the starvation period only cell death occurs. For a genotype $\alpha$ a fraction $\alpha$ of the cell population becomes spores and the remaining fraction, $1 - \alpha$, stays as vegetative cells. Spores are dormant cells and die at a constant rate $\delta$; vegetative cells die according to the survival curves detailed above, such that

$$\dot{X}_{\alpha,c}^{sp}(t) = -\delta X_{\alpha,c}^{sp}(t)$$

$$\dot{X}_{\alpha,c}(t) = -\beta(c)\mu(\mu t)^{\beta(c)-1}\left[X_{\alpha,c}(t) + \frac{\tilde{X}_{\alpha,c}e^{-(\mu T_{sur})^{\beta(c)}}}{1 - e^{-(\mu T_{sur})^{\beta(c)}}}\right] \quad (3)$$

where the second term in Eq. (3) fixes the cutoff in the life of vegetative cells and $\tilde{X}_{\alpha,c}$ is the initial population of the strain ($\alpha$, $c$) at the beginning of the starvation phase.

(2) Continuous (e.g. Phenotypic diversity). An alternative mechanism for spore production consists of vegetative cells becoming spores continuously at rate $\gamma$ (characteristic to the genotype) after the onset of starvation. In this case all cells have the potential to become spores, but some do so earlier than others. The final fraction of spores will be determined by the sporulation rate, the duration of the starvation phase, and the vegetative cell survival. The equations for the dynamics of both vegetative cells and spores have an additional term that reflects the transfer from the former to the latter,

$$\dot{X}_{\gamma,c}^{sp}(t) = -\delta X_{\gamma,c}^{sp}(t) + \gamma X_{\gamma,c}(t)$$

$$\dot{X}_{\gamma,c}(t) = -\beta(c)\mu(\mu t)^{\beta(c)-1}\left[X_{\gamma,c}(t) + \frac{\tilde{X}_{\gamma,c}e^{-(\mu T_{sur})^{\beta(c)}}}{1 - e^{-(\mu T_{sur})^{\beta(c)}}}\right] - \gamma X_{\gamma,c}(t) \quad (4)$$

*II. Simulation details.* We performed numerical simulations of growth-starvation cycles in environments defined by their starvation time (for deterministic ones) or by their average starvation time (for stochastic ones). For the latter, starvation times were a stochastic variable, exponentially distributed. The spectrum of genotypes was discretized, using different samplings depending on the spore production mechanism:

(1) Discrete mechanism. 44772 genotypes were used: 1092 values of $\alpha$ and 41 values of $c$. The values of $c$ range between $c = 0.05$ and $c = 0.45$ with a sampling of $10^{-2}$. The values of $\alpha$ were chosen with a sampling of $10^{-4}$ between $\alpha = 0$ and $\alpha = 0.1$ and with a sampling of $10^{-2}$ between $\alpha = 0.1$ and $\alpha = 1$. This irregular sampling was chosen to avoid abrupt jumps in the winning $c$ in environments that select for strategies with a small investment in spores.

(2) Continuous mechanism. The new terms in Eqs (4) corresponding to the formation of spores make them analytically intractable and computationally heavier. Thus only 99 genotypes were used: 11 values of $\gamma$ ($\gamma = 0$ and 10 additional values in the interval $[10^{-4}, 10^{-1}]$ given by $\gamma_i = 10^{(i-13)/3}$, equidistant in a logarithmic scale) and 9 values of $c$ evenly distributed in the interval $[0.05, 0.45]$.

Initial abundances of each genotype were independently drawn from a standard log-normal distribution and subsequently normalized so that the entire population contained $10^8$ cells. An initial resource pulse of magnitude $10^8$ was added and the trajectories governed by Eqs. (1) were integrated using finite-differences numerical methods until resources are fully consumed. Then, the starvation phase starts:

(1) Discrete mechanism. The starvation phase starts with the population splitting between spores and loners. A fraction $\alpha$ of cells turns into spores and the remaining $1 - \alpha$ remains as vegetative cells. Then, since Eqs (3) have an analytical solution, the functions are evaluated at $t = T$ (if the environment is stochastic, this time is previously sorted from an exponential distribution), which significantly speeds up the simulations.

(2) Continuous mechanism. Eqs (4) are integrated using finite-differences numerical methods until the end of the starvation period.

A new growth cycle starts with the arrival of a resources pulse of magnitude $10^8$. Then, the vegetative cells that survived the starvation period start reproducing according to Eq. (1), whereas the spores remain dormant for $\tau$ hours. After that time, spores germinate according to each genotype viability $v$. The spore viability was chosen to be a linear function of the reproduction rate $c$ (inversely proportional to cell size) and thus characteristic of each genotype:

$$v(c) = 1.1 - 2c \qquad (5)$$

Strains with the fastest reproduction have the lowest spore viability $v = 0.2$ and, at the other extreme, the strains with the slowest reproduction have the highest spore viability, $v = 1$. The fraction of spores resulting in viable cells was also multiplied by a parameter $s = 0.8$ to account either for unsuccessfully formed spores (asocial context) or for cells that are lost in the formation of an aggregate (e.g. stalk formation in the social amoeba *D. discoideum*).

To determine the winning genotype, in deterministic environments a single run was used while in stochastic environments averages were taken over 20 independent realizations. In both cases simulations were run until $t = 10^8$ hours and the winning genotype defined as the most abundant at that time. In stochastic environments the variance of this measurement is very low and the mean value coincides with the result of each single run. Figures S1 and S2 show the short time evolution of some genotypes in deterministic and stochastic environments.

*Chimeric success.* Genotypes underwent one growth-starvation cycle. Let $X^{sp}_{*,c}$ be the amount of spores produced by genotype ($*$, $c$) instantaneously upon starvation, where the $*$ is a placeholder for either $\alpha$ or $\gamma$. The pairwise chimeric success of genotype ($*$, $c$) against genotype ($*'$, $c'$) is given by $X^{sp}_{*,c}/(X^{sp}_{*,c} + X^{sp}_{*',c'})$ and the average chimeric success of genotype ($*$, $c$) is given by:

$$CS(*,c) = \sum_{*',c'} \frac{X^{sp}_{*,c}}{X^{sp}_{*,c} + X^{sp}_{*',c'}} \qquad (6)$$

where the sum runs over all the possible values of $*$ and $c$ except for $* = *'$ and $c = c'$ simultaneously.

*Linear hierarchies of chimeric dominance* (Fig. 5). To construct the linear hierarchies the genotypes are ordered according to the environment they evolved in. Label 0 corresponds to the winning genotype from the environment with $\lambda_T = 10$ hour and label 30 to the winning genotype that evolved in the environment with $\lambda_T = 10^4$ hour. All strains are mixed in pairs, they undergo a growth-starvation cycle, and pairwise chimeric success is measured. To define a dominance / subordination relationship between the genotypes we set one of the genotypes in the mix as the reference genotype and its partner as the mixed genotype. If the reference genotype has a higher chimeric success the encounter will be dominant; if it has a smaller chimeric success the encounter will be subordinate; and it will be neutral if both make the same number of spores.

*Development time* (Fig.2). To obtain the speed with which a genotype $\gamma$ completes its sporulation, clonal populations of the same initial size are initialized without food and allowed to grow until they starve. In response to starvation, vegetative cells turn into spores at a rate $\gamma$. Spores die at rate $\delta$. Vegetative cells die according to the survival curve of their respective genotype. The spore formation dynamics stops after time $T_{sur}$, when all the vegetative cells have either died or have become spores. Finally, the number of spores at every time is normalized by the final number of spores to obtain the rate at which this maximum value is reached.

**Acknowledgments**: We thank: the IFISC (CSIC-UIB) computing lab for technical support and the use of their computational resources; and A. Herre for helpful discussions.

**Supporting information captions.**

| Description | Parameter | Value | Units |
|---|---|---|---|
| Rate of change of the vegetative cell death rate | $\mu$ | $2 \times 10^{-3}$ | hour$^{-1}$ |
| Maximum lifetime of a vegetative cell | $T_{sur}$ | 200 | hour |
| Spore germination time | $\tau$ | 4 | hour |
| Spore mortality rate | $\delta$ | $2 \times 10^{-4}$ | hour$^{-1}$ |
| Fraction of cells that successfully completes sporulation | $s$ | 0.8 | --- |
| Initial food pulse | $R_0$ | $10^8$ | # cells |
| Half-saturation constant of resources consumption | $R_{1/2}$ | $0.1 R_0$ | # cells |
| Division rate | $c$ | Varied | hour$^{-1}$ |
| Investment in spores (discrete mechanism) | $\alpha$ | Varied | --- |
| Sporulation rate (continuous mechanism) | $\gamma$ | Varied | hour$^{-1}$ |

**Table S1: Description of parameters and values used.** The lower part of the table gives the parameters that are allowed to evolve.

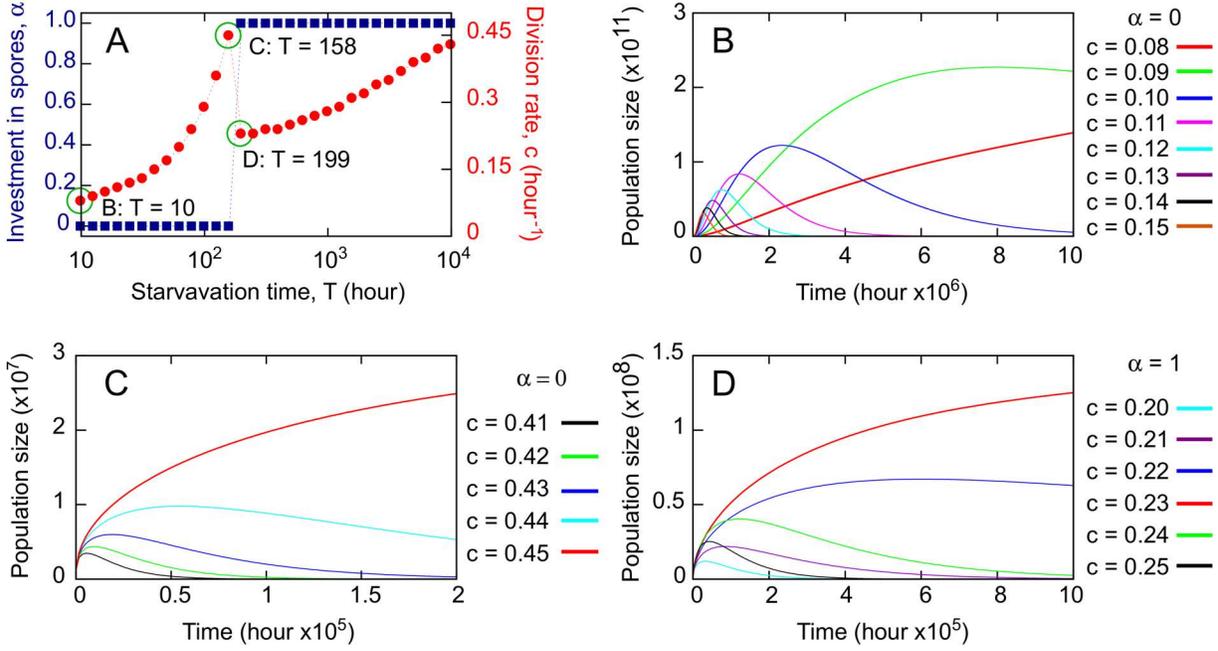

**Figure S1: Short-time evolution of populations in deterministic environments.** A) Winning genotype in each environment, given by its investment in spores $\alpha$ (blue squares), and division rate $c$ (red circles). Three environments are chosen and marked with a green circle, two of them select for $\alpha = 0$ strategies and one for $\alpha = 1$. B) Fast-recovery environment ($T = 10$ hour): fast reproducing strains have an advantage in the short run, but are in the long run outcompeted by the genotype with the lowest division rate. C) Intermediate-recovery environments ($T = 158$ hour): fast-reproducing strains are able to maintain their competitive advantage throughout. D) Slow-recovery environments ($T = 199$ hour): selection for genotypes with slow cell division.

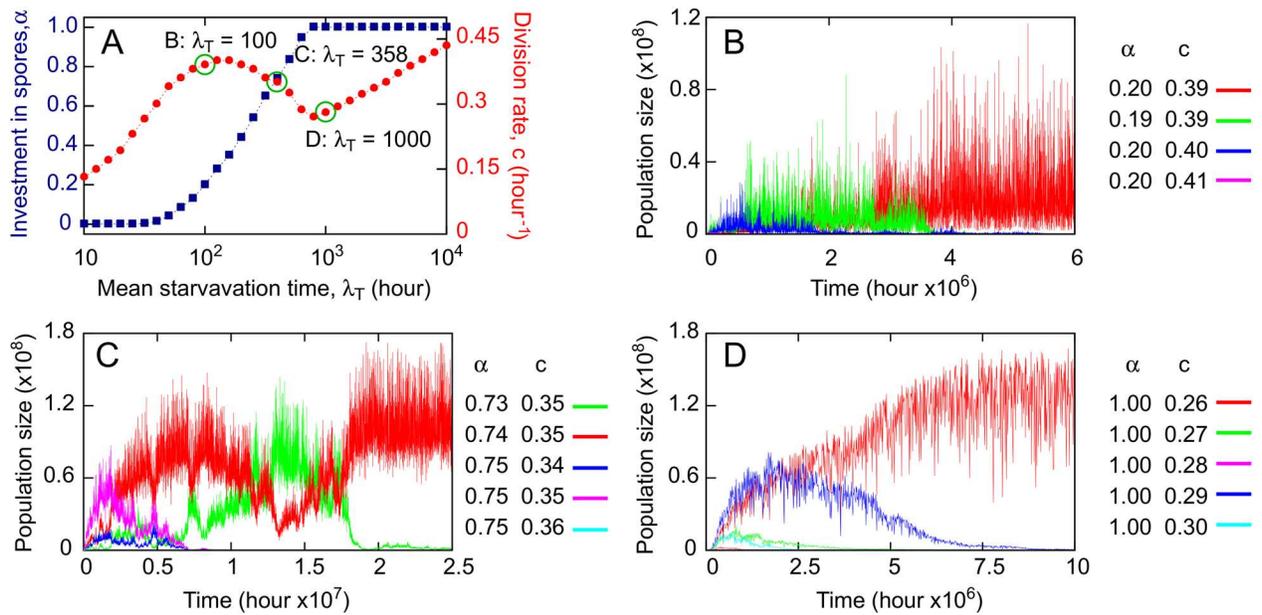

**Figure S2: Temporal evolution of populations in stochastic environments.** A) Winning genotype in each environment, given by its investment in spores, $\alpha$ (blue squares), and division rate, $c$ (red circles). Three environments are chosen and marked with a green circle. B,C) For intermediate environments ($\lambda_T = 100$ hour and $\lambda_T = 358$ hour respectively), populations show high amplitude fluctuations since the variation in the starvation times favors in each cycle a different genotype. D) For slow environments ($\lambda_T = 1000$ hour), the amplitude of the fluctuations in the population size decreases since most of the starvation times favor genotypes with $\alpha = 1$.

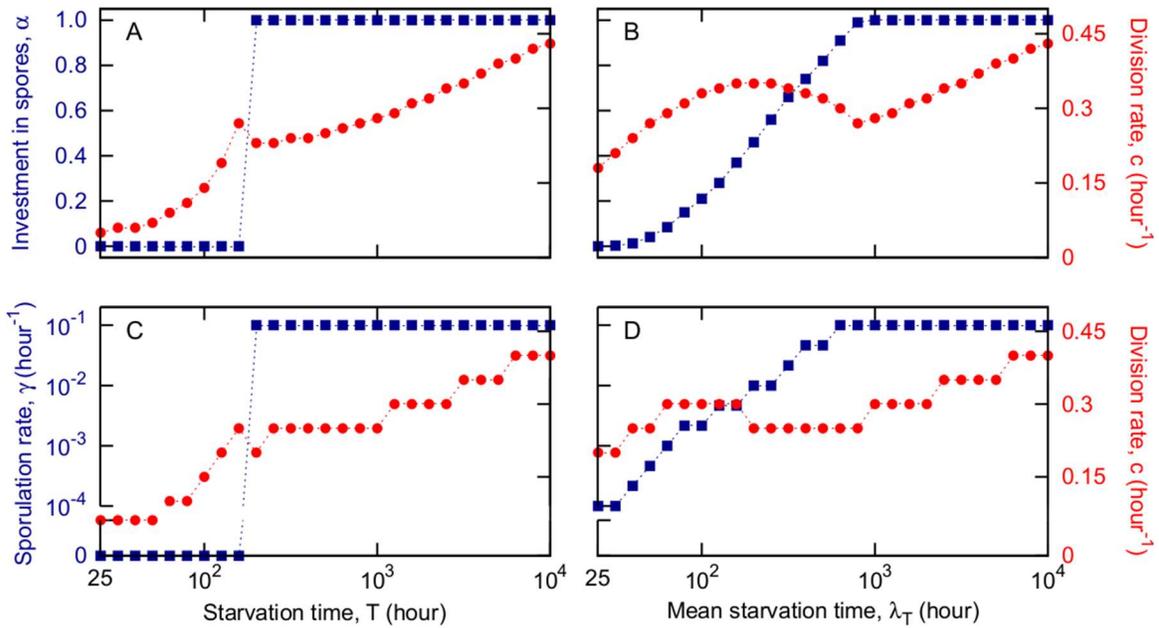

**Figure S3: Sensitivity analysis of the winning genotype for the choice of survival cost function, $\beta$.** Increasing the survival probability of bigger cells over time ($\beta = 5.1 - 8c$) reduces selection for fast reproducing strains. The cost of small spores in terms of germination survival was not changed, so fast spore-selecting environments select for the same division rate as in Fig. 3. A) Discrete spore-making and deterministic environments. Winning genotype was determined as the most abundant out of 4141 (101 values of $\alpha$ and 41 values of $c$) after a single realization of $10^8$ hour. B) Discrete spore-making and stochastic environments. Results averaged over 20 independent simulation runs starting with 44772 genotypes (1092 values of $\alpha$ and 41 values of $c$). C) Continuous spore-making and deterministic environments. Winner obtained as in A from a single realization starting from 99 genotypes (9 values of $c$ and 11 values of $\gamma$). D) Continuous spore-making and stochastic environments. Setup as in C but with exponentially distributed starvation times; averages taken over 20 independent realizations.

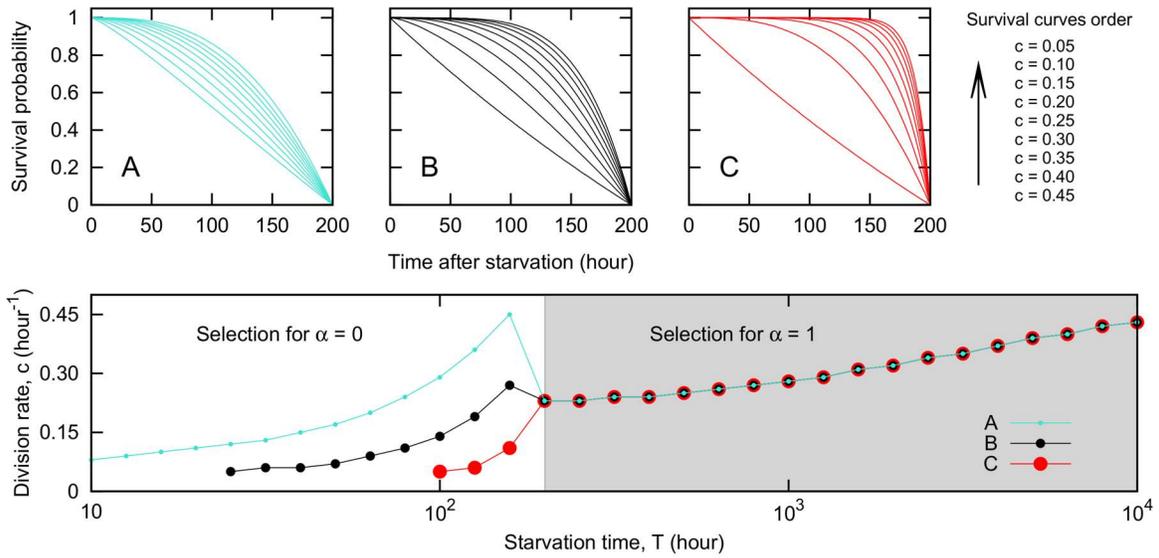

**Figure S4: Sensitivity analysis for different choices of survival cost function $\beta$ in deterministic environments.** Modifying the tradeoff between cell survival and cell size leads to selection for different division rates in the $\alpha = 0$ environments. The top row shows three families of curves where the survival advantage of bigger cells against the smaller ones increases from left to right. A) $\beta = 3.1 - 4c$, B) $\beta = 5.5 - 10c$, C) $\beta = 19 - 40c$. D) Selected division rate as a function of the starvation time in deterministic environments.

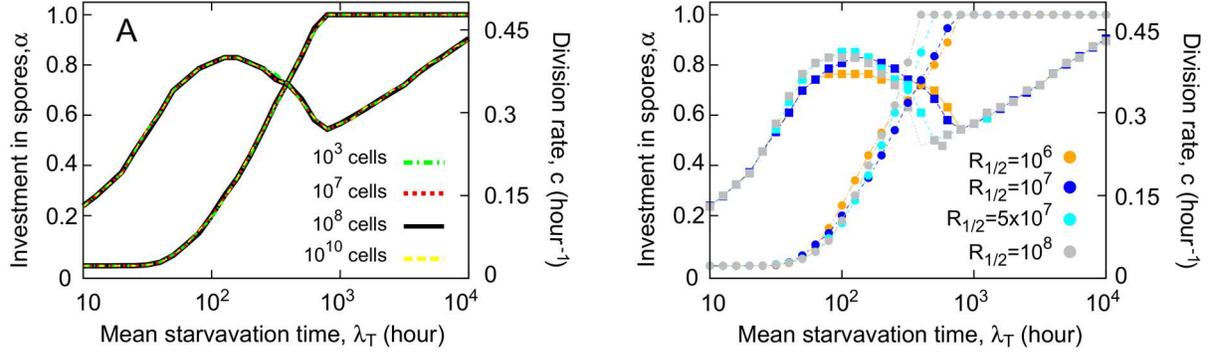

**Figure S5: Sensitivity analysis for the initial condition and the Michaelis-Menten rates.** In both panels simulations were performed for the discrete spore-forming mechanism in stochastic environments and starting with 44772 genotypes (1092 values of $\alpha$ and 41 values of $c$). A) Initial population size does not modify the winning genotype, only the transient dynamics. Highly diluted initial populations initially favor genotypes with a high division rate. However once the total population reaches the carrying capacity genotypes with fast division start declining and eventually go to extinction. On the contrary, if the initial population is above the carrying capacity the total number of cells decreases at short times. Bigger cells take advantage of their longer survival but as the population reaches the carrying capacity, genotypes with a higher division rate start growing and finally outcompete the slower strains. Initial populations of each genotype were drawn from a log-normal distribution and the total population subsequently normalized to $10^3$, $10^7$, $10^8$ and $10^{10}$ cells. The size of the food pulse was kept constant, so increasing the population size increases the competition for resources. Data points are not shown for clarity, sampling in the mean starvation time as in panel B. B) Increasing the saturation constant $R_{1/2}$ anticipates selection for pure sporulating strategies ($\alpha = 1$) since the growth term decreases. Environments with a given mean starvation time become harsher and it is more beneficial to make more spores and reproduce faster. The initial total population was fixed at $10^8$ cells and the amplitude of each food pulse at $10^8$. Squares indicate division rate and circles investment in spores.

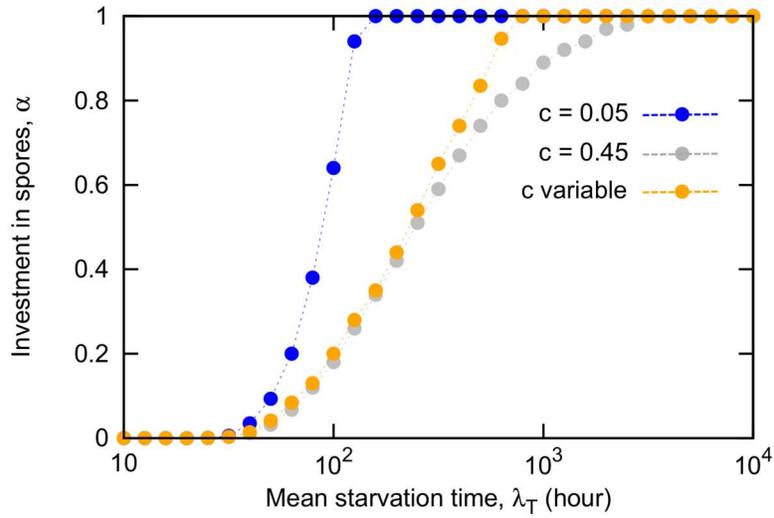

**Figure S6: Interaction between spore investment and division rate.** Fixing the division rate modifies the optimal value of the investment in spores. Simulations with variable $c$ were started with 44772 genotypes (1092 values of $\alpha$ and 41 values of $c$). When the division rate is fixed as a parameter, only the 1092 genotypes corresponding to $\alpha$ were taken.

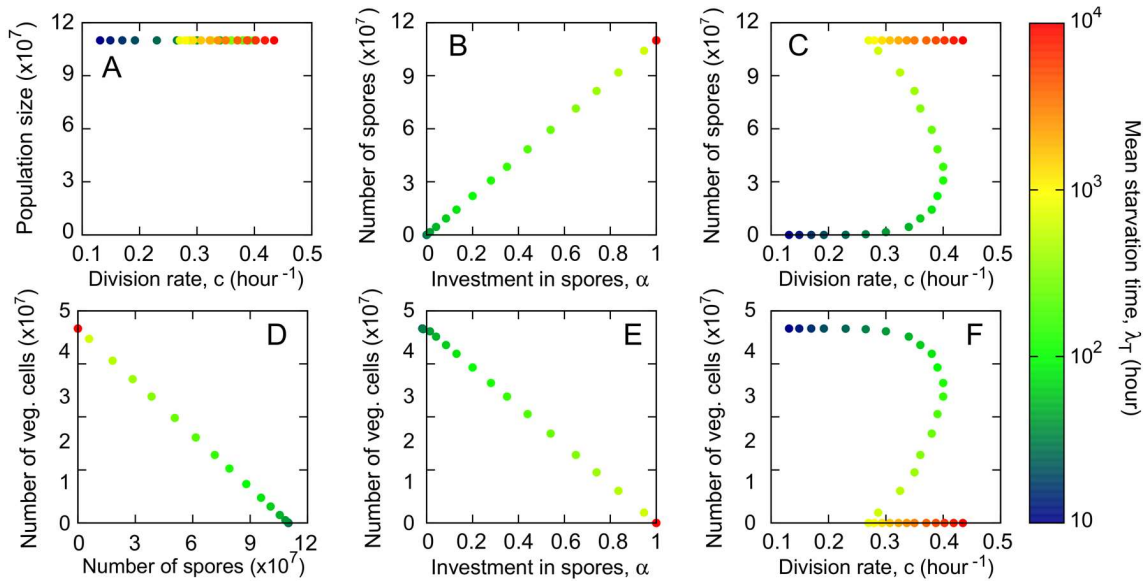

**Figure S7: Life history tradeoffs between non-social traits in genotypes with discrete spore formation.** Different traits are measured immediately after the population divides between spores and vegetative cells. A) The total population is constant for all genotypes. Number of spores versus B) the spore investment, $\alpha$; C) division rate, $c$. Number of vegetative cells versus D) the number of spores; E) the investment in spores; F) the division rate.

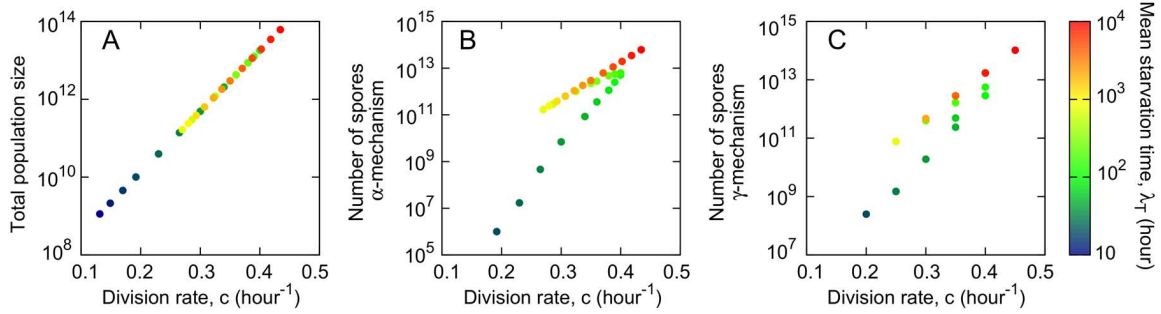

**Figure S8: Tradeoff between number of spores and cell and spore size under different experimental settings**. When the strains are plated on abundant resources and grow exponentially during a fixed time, followed by sudden starvation, A) the population size, B) the number of spores (discrete mechanism) and C) the number of spores (continuous mechanism) all correlate positively with the division rate. Logarithmic scale used for the vertical axis in all the panels.

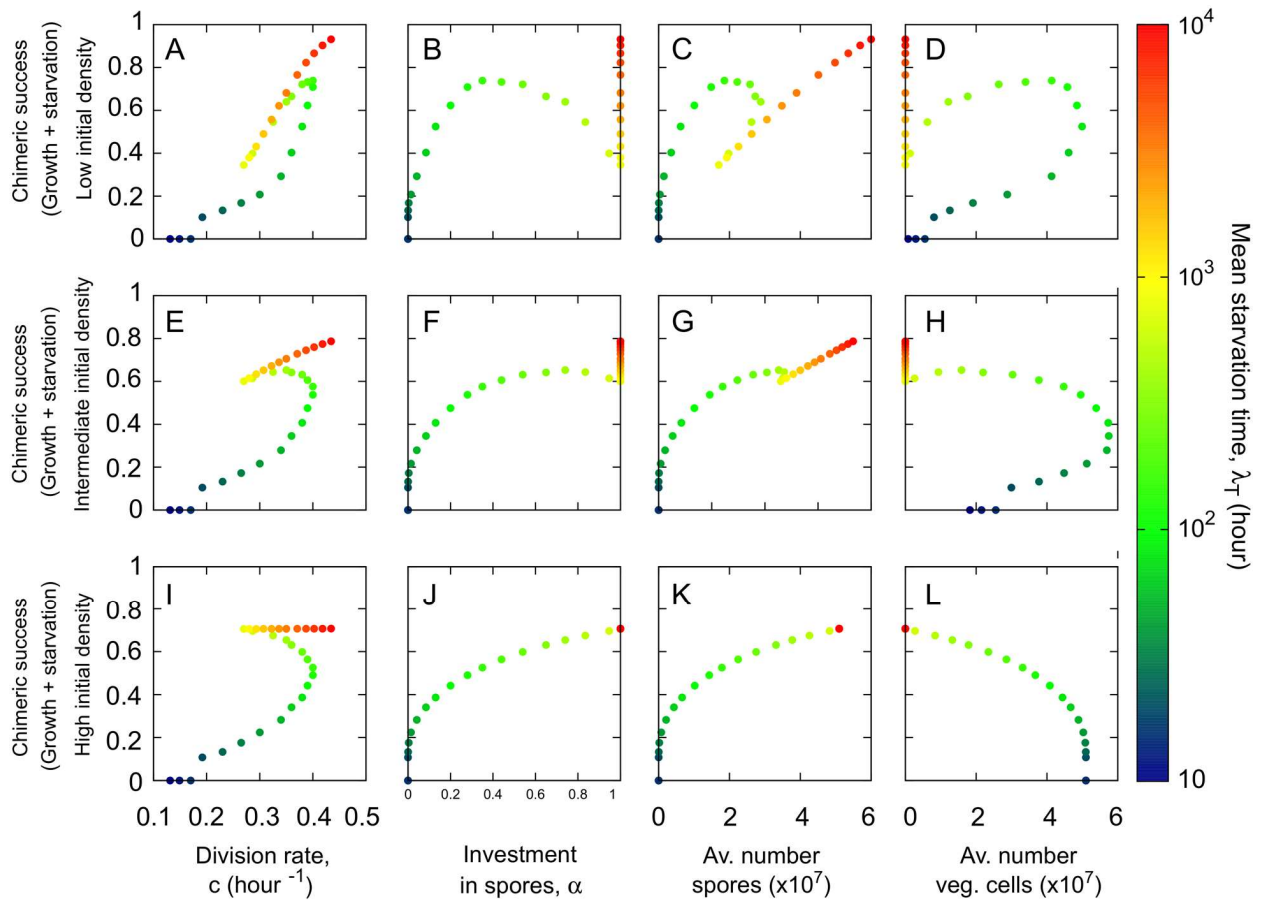

**Figure S9: Apparent life history tradeoffs involving social traits in genotypes with discrete formation of spores.** Including the growing phase in the measurement of chimeric success modifies the results, equivalently to the continuous spore formation mechanism. From left to right: chimeric success versus division rate (A, E, I), investment in spores (B, F, J), average number of spores (C, G, K) and average number of vegetative cells (D, H, I). From top to bottom: low cell:resource initial density ($10^3$ cells and $R_0 = 10^8$), intermediate cell:resource initial density ($10^7$ cells and $R_0 = 10^8$) and high cell:resource initial density ($10^{10}$ cells and $R_0 = 10^8$).